\documentclass[aps,pre,twocolumn]{revtex4-1}
\usepackage{epsfig}
\usepackage{natbib}
\usepackage{ulem}
\usepackage{xspace}
\usepackage{bbold}
\usepackage{appendix}
\usepackage{natbib}
\usepackage{graphicx}
\usepackage{amsmath}
\usepackage{amssymb}
\usepackage[dvipsnames]{xcolor}
 
\begin{document}
\title{Analytical results for a minimalist  thermal diode}

\author{Lucianno  Defaveri$^1$}
\author{Celia Anteneodo$^{1,2}$}
 \affiliation{$^1$Department of Physics, PUC-Rio, Rio de Janeiro, 22453-900 RJ, Brazil}
\affiliation{$^2$Institute of Science and Technology for Complex Systems, Rio de Janeiro, Brazil}

\begin{abstract}
We consider a system consisting of two interacting classical particles, 
each one subject to an on-site potential and to a Langevin thermal bath. 
We analytically calculate the heat current that can be established through the system when the bath temperatures are different, for weak nonlinear forces.  
We explore the  conditions under which the diode effect emerges when inverting the temperature difference. 
Despite the simplicity of this two-particle diode, an  intricate dependence on the system parameters is put in evidence. Moreover, behaviors reported for long chains of particles can be extracted,  
for instance, the dependence of the flux with the interfacial stiffness and type of forces present, as well as the dependencies on the temperature required for rectification. These analytical results can be a tool to foresee the distinct role that diverse types of nonlinearity and asymmetry  play in  thermal conduction and rectification. 
\end{abstract}

\maketitle

\section{Introduction}
 
Simplified microscopic models, such as classical particle chains in contact with heat baths, have proven useful to grasp  the physics of thermal transport in low dimensions~\cite{fourier,ReviewLepriLiviPoliti2003,ReviewDhar2008,LepriBook2016}. 
The interest in one-dimensional models goes beyond  the  theoretical challenge to derive the laws of heat conduction from the microscopic dynamics, insofar as they can be useful for  understanding the  anomalies  observed in real systems, such as carbon nanotubes~\cite{nano2008}, nanowires~\cite{YangZhangLi2010}, and molecular chains~\cite{chain1,chain2}.
Moreover, these  experiments and theories can
lead to the development of new technologies for heat flow manipulation~\cite{e1}. An interesting example  is the thermal diode, whose thermal conductivity along a given axis changes depending on the direction of the heat flux, yielding rectification in a preferential direction. 
This proposal, initially conceived through  simplified theoretical modeling
~\cite{casati-diode}, soon found   materialization in  solid-state experiments~\cite{exp-diode}.

Subsequently, several variants of microscopic models were proposed to determine the conditions to achieve efficient  rectification, 
by analyzing for instance the effects of the range of the interactions or graded masses~\cite{efficiency1,efficiency2}, the role of the interface~\cite{interface,pons2017,efficiency3,BaowenLi2005,BaowenLi2011}, among others. 
In the meantime, 
several efforts have been directed towards an analytical  
understanding of the diode effect,  
putting into evidence the requirements of asymmetry and nonlinearity for rectification, for instance, by linearizing the equations of motion but, as  counterpart, making the parameters along a mass graded chain temperature dependent~\cite{Tdependence}. Closely related, in a very recent work, the diode effect was shown in the so-called temperature-gradient harmonic oscillator chains~\cite{harmonic}. 
In the same spirit, a minimalistic model of two harmonic oscillators with temperature dependency has been recently studied~\cite{Muga2021}.  
The two-segment chain of classical
spins in contact with multiple heat baths
has also been studied~\cite{spins}, 
as well as quantum systems, 
to show rectification of the heat flow between two thermal
baths through a pair of interacting qubits~\cite{quantum-diode}
or even quantum spin chains~\cite{perfectdiode}.

Here, we investigate a minimalist model of only two interacting classical particles 
connected to heat baths, in order to understand the diode effect directly from the equations of motion. %
We solve these equations in the limit of small nonlinearity, from a perturbative approach. Then, we obtain expressions for the heat flow and rectification factor allowing us to directly grasp the impact of  asymmetries and nonlinearities, as well as qualitative features of heat conduction and rectification, 
explicitly expressed in terms of the model parameters. 

The paper is organized as follows. The system is defined in Sec.~\ref{sec:system}. The perturbative solution and associated heat flow are described in Secs.~\ref{sec:solution} and \ref{sec:heatflow}, respectively, while the mathematical derivations can be found in the Appendix.  
The diode effect is discussed in Sec.~\ref{sec:rectification} 
with final remarks in Sec.~\ref{sec:final}.

\begin{figure}[b!]
\centering 
\includegraphics[width = 0.35\textwidth]{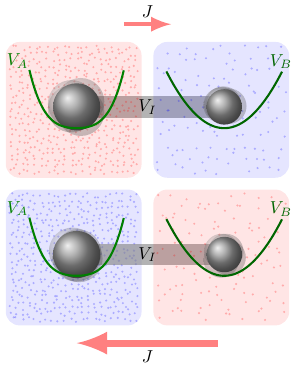}
\caption{Schematic representation of a minimalist  thermal diode.}
\label{fig:two-particle-model} 
\end{figure}

\section{The system}
\label{sec:system}

We consider a one-dimensional system composed of two particles,  with masses $m_j$ (with $j=A,B$), coordinates $x$ and $y$,  
subject to on-site potentials $V_j$  and interacting through a potential $V_I$ such that the complete Hamiltonian is 
\begin{eqnarray}
\mathcal{H}
= \frac{p_A^2}{2m_A}  + V_A(x) +\frac{p_B^2}{2 m_B} + V_B(y) +  V_I (x-y).
\end{eqnarray}
Moreover, each particle $j$ is put in contact with a Langevin thermal bath at temperature $T_j$. 
A pictorial representation of this kind of system is provided in Fig.~\ref{fig:two-particle-model}. 
Let us remark that this system is very similar to the couple of harmonic oscillators recently investigated~\cite{Muga2021}, but in our case we introduce nonlinear forces. 
Namely, 
we treat the case where the interaction potential is harmonic with stiffness $k_I$, while the 
 on-site potential of particle $j$ is  $V_j(z)= k_j z^2/2 + \epsilon V^{nl}_j(z)$, where $k_j$ is  the harmonic stiffness and $f_j(z)= -d V^{nl}_j(z)/dz$ is an arbitrary nonlinear force, whose intensity is 
 controlled by the unitless constant  $\epsilon$.  
 Explicitly, the equations of motion are
\begin{eqnarray} \label{eq:motionx}
	m_A \ddot{x} + \gamma_A \dot x + k_A x + k_I (x - y) &=& \epsilon f_A(x) + \eta_A(t), \\ \label{eq:motiony}
	m_B \ddot{y} + \gamma_B \dot y + k_B y + k_I (y - x) &=& \epsilon f_B(y) + \eta_B(t) \, ,
\end{eqnarray}
 where $\epsilon$  is a dimensionless parameter that  controls the strength of the nonlinear forces, 
$\gamma_j$ is the damping coefficient and $\eta_j$ the fluctuating force of the Langevin thermostat $j$ ($j=A,B$), 
where $\eta_A$ and $\eta_B$  are independent zero-mean Gaussian-distributed white noises with
\begin{eqnarray} \nonumber
\langle \eta_A(t)\eta_A(t') \rangle &=& 2 \gamma_A T_A \delta(t-t'), \\ \nonumber
\langle \eta_B(t)\eta_B(t') \rangle &=& 2 \gamma_B T_B \delta(t-t') \, ,  
\end{eqnarray}
where the temperature is in units of the Boltzmann constant.
Although we might suppress some parameters by fixing space and timescales, we will keep them explicit  to preserve the 
AB symmetry of the equations.

\section{Perturbative solution}
\label{sec:solution}

Eqs.~(\ref{eq:motionx})-(\ref{eq:motiony}) cannot be solved exactly, however, if $\epsilon$ is small enough to ensure that the energy stored in the nonlinear mode is much smaller than in the harmonic one, we can expand the coordinates  as 
\begin{eqnarray} 
x(t) &=& x_0(t) + \epsilon \, x_1(t) + \mathcal{O}(\epsilon^2), \label{eq:expansionx} \\
y(t) &=& y_0(t) + \epsilon \, y_1(t) + \mathcal{O}(\epsilon^2) , \label{eq:expansiony}
\end{eqnarray}
where the  zeroth-order terms follow the equations 
\begin{eqnarray}
m_A\ddot{x}_0 + \gamma_A \dot{x}_0 + k_A x_0 + k_I (x_0 - y_0) &=& \eta_A(t), \label{eq:x0}  \\
m_B\ddot{y}_0 + \gamma_B \dot{y}_0 + k_B y_0 + k_I (y_0 - x_0) &=& \eta_B(t),
\label{eq:y0}
\end{eqnarray}
which are linear and uncoupled equations, 
and the first-order corrections $x_1$ and $y_1$ follow 
\begin{eqnarray}
m_A \ddot{x}_1 + \gamma_A \dot{x}_1 + k_A x_1  + k_I (x_1 - y_1) &=& f_A(x_0) \,, \label{eq:x1} \\
m_B \ddot{y}_1 + \gamma_B \dot{y}_1 + k_B  y_1  + k_I (y_1 - x_1) &=& f_B(y_0). \label{eq:y1}
\end{eqnarray}

Since we are interested in the long-time behavior, the initial conditions are not relevant, therefore we will use the Fourier transform, defined as
$ \tilde{z}(\omega) = \int_{-\infty}^{\infty} dt  \, z(t) \, {  e}^{-i \omega t}$, to solve the above  stochastic differential equations. 
We start by expressing Eqs.~(\ref{eq:x0})-(\ref{eq:y0}) in Fourier space, namely,
\begin{eqnarray} \label{eq:0x}
	\underbrace{\big( k_A + k_I - m_A\omega^2 + i \gamma_A \omega \big)}_{ \displaystyle a(\omega)} \tilde  x_0 - k_I \tilde y_0   &=& \tilde \eta_A(\omega) \,, \\  \label{eq:0y}
	\underbrace{\big( k_B + k_I - m_B\omega^2 + i \gamma_B \omega \big)}_{\displaystyle  b(\omega)} \tilde  y_0 - k_I \tilde x_0 &=& \tilde \eta_B(\omega) \, ,
\end{eqnarray}
whose solution in matrix form is 
\begin{eqnarray} \label{eq:xyw0}
\begin{pmatrix}
\tilde{x}_0(\omega) \\
\tilde{y}_0 (\omega)
\end{pmatrix} =\frac{ 
\begin{pmatrix}
b(\omega) & k_I\\
k_I & a(\omega)
\end{pmatrix}
\begin{pmatrix}
\tilde{\eta}_A(\omega) \\
\tilde{\eta}_B (\omega)
\end{pmatrix}
}{ a(\omega)b(\omega)-k_I^2 }\,. 
\end{eqnarray}
Similarly, solving Eqs.~(\ref{eq:x1}) and (\ref{eq:y1}) in Fourier space,  
we obtain (for more details see \ref{app:xy})

\begin{eqnarray} \label{eq:xyw1}
\begin{pmatrix}
\tilde{x}_1(\omega) \\
\tilde{y}_1 (\omega)
\end{pmatrix}=\frac{ 
\begin{pmatrix}
b(\omega) & k_I\\
k_I & a(\omega)
\end{pmatrix}
\begin{pmatrix}
\mathcal{F} \left\{  f_A({x}_0) \right\} (\omega) \\
\mathcal{F} \left\{  f_B({y}_0) \right\} (\omega)
\end{pmatrix}
}{ a(\omega)b(\omega)-k_I^2 }\,. 
\end{eqnarray}

\section{Heat flow}
\label{sec:heatflow}

The heat flow $J$ along the system can be defined in several forms that are equivalent when the system reaches a stationary state (see for instance \cite{ReviewDhar2008}). 
The potential $V_I(x-y)$ represents the energy stored in the interaction between neighboring particles, and the energetic flow can be written as
\begin{eqnarray}
\frac{d}{dt} \big\langle V_I(x-y) \big\rangle 
&=&  \big\langle V'_I(x-y) \dot{x} \big\rangle -   \big\langle V'_I(x-y) \dot{y} \big\rangle , \nonumber
 \end{eqnarray}
and, under stationarity,   
$\langle V'_I(x-y) \dot{x} \rangle =   \langle V'_I(x-y) \dot{y}  \rangle$. From this identity, we have   equivalent   definitions that in our case, where $V_I'(x-y) = k_I (x-y)$, read
\begin{eqnarray}  \label{eq:Jdef}
 J &=&  \langle k_I(x-y) \dot{x}  \rangle  =   \langle k_I(x-y) \dot{y}  \rangle\,,  
 \end{eqnarray}
 or still $J=\langle k_I(x-y) (\dot{x} + \dot{y})/2  \rangle$.
The heat flow $J$ can also be expanded in a series of $\epsilon$, as
\begin{eqnarray} \label{eq:J0eJ1}
J &=& J_0 + \epsilon J_1 + O(\epsilon^2) \, .
\end{eqnarray}
In the linear regime,   Eq.~(\ref{eq:Jdef}) yields (see Appendix \ref{app:heat})
\begin{eqnarray} \label{eq:Jx}
J_0 &=&	 k_I \langle  x_0 \dot{x}_0 \rangle - k_I \langle  y_0 \dot{x}_0 \rangle \\
&=& -  k_I \int \frac{d \omega d \omega'}{(2 \pi)^2} e^{i t (\omega + \omega')}   \langle  \tilde y_0 (\omega) \; i \omega' \;\tilde{x}_0(\omega') \rangle \, . \;\;\;
\end{eqnarray}  
Then, we obtain
\begin{eqnarray} \label{eq:J0}
	J_0	&=&   \kappa_0  \big(T_A - T_B\big)   \, ,
\end{eqnarray}
where the zeroth-order thermal conductivity is  
\begin{eqnarray} \label{eq:kappa0}
	\kappa_0 &=&  \int \frac{d \omega}{2 \pi}  
	\frac{2 \gamma^2 k_I^2 \omega^2 }{|a(\omega)b(\omega) - k_I^2|^2}  \,.
\end{eqnarray}
{\bf Note:} This expression shows that,  regardless of the asymmetries that may be present, the linear system cannot be converted to a thermal diode since $\kappa_0$  does not depend on the temperatures  and it is invariant under particle exchange, hence, the magnitude of the flow is the same in both directions.

With regard to the dependence of 
$\kappa_0$ on the coupling strength $k_I$,  Eq.~(\ref{eq:kappa0}) gives 
	$\kappa_0 \sim  k_I^2  + O(k_I^3)$, 
for small $k_I$. It is interesting to note that this is the scaling observed for  two-segment chains with nonlinear forces of Frenkel-Kontorova (FK) type~\cite{casati-diode}. 

Now we proceed to calculate the first-order 
correction of the current $J$, containing the information on the nonlinearities.  
From Eq.~(\ref{eq:Jdef}), we have 
\begin{eqnarray} \nonumber
	J &=& - k_I \langle y(t) \dot{x}(t) \rangle 
	=\\ \label{eq:J01}
	&=& 
	\underbrace{- k_I \langle y_0 \dot{x}_0 \rangle}_{\displaystyle J_0}  + \epsilon \underbrace{ (-k_I)   \{  \langle y_1 \dot{x}_0 \rangle + \langle y_0 \dot{x}_1 \rangle  \} }_{\displaystyle  J_1}  \, ,
\end{eqnarray}
where  the correlations in $J_1$ are calculated in Appendix~\ref{app:heat}, yielding $J_1=\kappa_1(T_A,T_B)\,\Delta T$, with
\begin{eqnarray} 
\kappa_1(T_A,T_B) &=&  
\kappa_0 \sum_{j=A,B} \beta_j \,   g_j\big(\sigma_{m j} T_m  +  \sigma_j T_j \big), \;\;\;\;\;\; \label{eq:kappa1}
\end{eqnarray}
where $T_m=(T_A+T_B)/2$ is the mean temperature, $\beta_j$, $\sigma_{m j}$
  and $\sigma_j$ ($j=A,B$) are coefficients that do not depend on the temperature (derived in  Appendix \ref{app:heat}, where explicit expressions are also given),  
and $g_j$ is  derived from the Fourier transform of the force $f_j$ (for examples see Table I).

\begin{table}[h!]
\begin{tabular}{ |c|c| } 
 \hline
$  f_j(z) = \sum_{n\ge 0}   c_{j,n}   z^n$ & $ g_j(z)= \sum_{n\ge 1} c_{j,n}\,  n!!\,   z^{(n-1)/2}$  \\[3mm]   
    \hline
    \hline
    $ -z^{2n-1}$, $n\in \mathbb{N}$ &  $ -(2n-1)!!\,z^{n-1}$   \\[2mm]  
    \hline
$-\sin(Kz)$  & $ -K{ e}^{-K^2z/2}$  \\[2mm] 
   \hline
$-\sinh(Kz)$ & $ -K{ e}^{ K^2 z/2}$  \\[2mm]  
   \hline
\end{tabular}
\caption{Nonlinear force $f_j$ and associated function $g_j$ (derived in Appendix~\ref{app:heat}). 
 }
\end{table}

In Fig.~\ref{fig:num},  an illustrative example shows the good agreement between the first-order theoretical prediction, Eq.~(\ref{eq:J01}), and 
the numerical evaluation of Eq.~(\ref{eq:Jdef}), 
performed over trajectories obtained from the numerical integration of the equations of motion, using an eighth-order Runge-Kutta algorithm~\cite{RK8}.

 \begin{figure}[h!]
\centering 
\includegraphics[width = 0.5\textwidth]{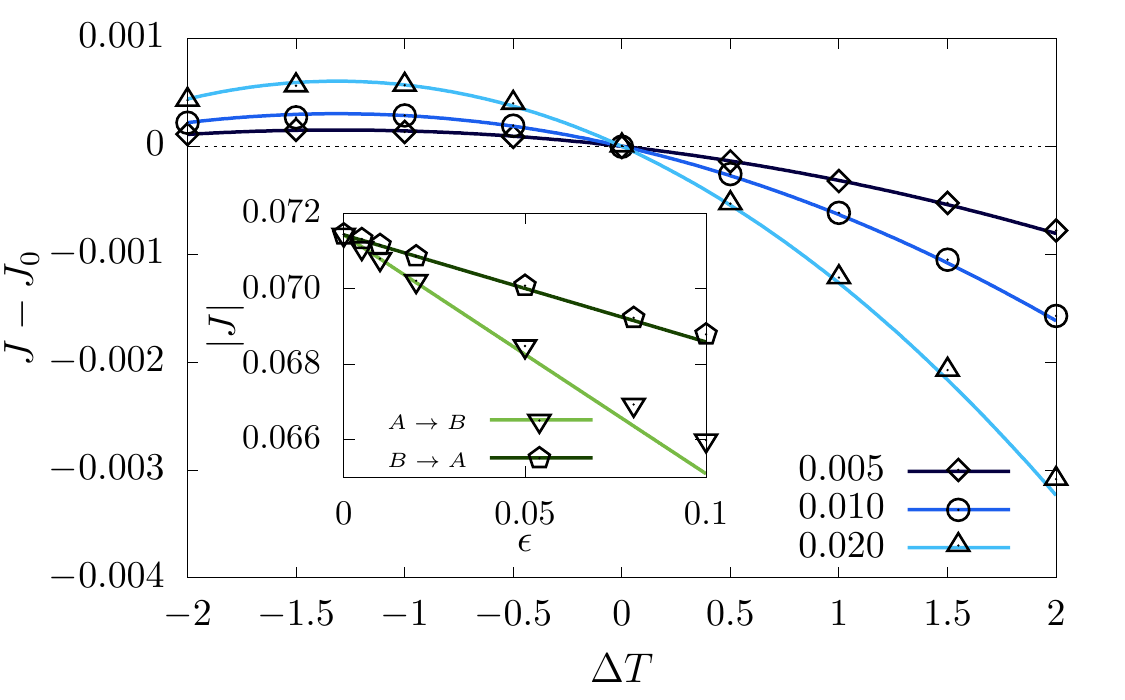}
\caption{  Heat current difference $J-J_0$ vs. $\Delta T$. Solid lines correspond to the theoretical prediction given by Eq.~(\ref{eq:J01}) and symbols to the computation from numerical integration of the equations of motion, averaged over $10^5$ realizations.  
$V_A^{nl}(z)=z^4/4$, $V_B^{nl}(z)=0$, 
Different values of $\epsilon$ indicated on the figure were considered, $k_I=0.5$ and all other parameters are equal to 1. The inset shows the current $|J|$ as a function of $\epsilon$ for $\Delta T=\pm 1$ (i.e, $A\rightleftarrows B$). 
(Color online.)
}
\label{fig:num} 
\end{figure} 
 
For weak coupling, the scaling
\begin{eqnarray}
	\kappa \sim  k_I^2 + O(k_I^3,\epsilon k_I^2) \, ,
\end{eqnarray} 
obtained in the linear case, still holds, while  $\kappa$ tends to a constant value for large $k_I$. 
Of course, if the interfacial interaction, which connects the two units of the system,   vanishes, hence, $\kappa$ vanishes too, as expected due to disruption of  the channel for energy flux.  
  
As a check of consistency, 
we verified  that,
if $f_j(z)$ were linear, then $g_j(z)$ would be a constant $\bar{g}_j<0$ ($n=0$ in Table I), 
in which case Eq.~(\ref{eq:kappa1}) becomes
$\kappa_1 =(\beta_A \bar g_A+\beta_B \bar g_B)\,\kappa_0$\,
in accord with the zeroth order expression 
for $\kappa_0$, Eq.~(\ref{eq:kappa0}),  after substituting 
$k_A \to k_A - \epsilon \bar{g}_A$ and 
$k_B \to k_B - \epsilon \bar{g}_B$.     

Since the linear conductivity $\kappa_0$ does not depend  on the bath temperatures, the heat flux will have the same magnitude in both directions.
Therefore, for rectification, it is crucial that  at least one of the two forces $f_j(z)$ be nonlinear. 
This nonlinearity would act by introducing a dependence of the conductivity on the temperatures, through the argument of $g_j(z)$ in   Eq.~(\ref{eq:kappa1}), which originates from the correlations of the zeroth-order coordinates.
A temperature dependence that is asymmetric under particle exchange  is then responsible for thermal rectification, as   discussed in the next section. 
Moreover, this is the basis of the diode effect on harmonic systems  with imposed or natural temperature dependencies~\cite{Tdependence,harmonic,Muga2021}.

\section{Diode effect}
\label{sec:rectification}

First, recall that $\beta_j$, $\sigma_{m j}$ and $\sigma_j$, which define $J_1$, are quantities that do not depend on the end temperatures.  
While  $\sigma_{m j}$ (as well as $\kappa_0$) is always positive, $\beta_j$ and $\sigma_j$ have not definite sign in general, however, $\sigma_{m j}T_m+\sigma_j T_j$ must be positive. 
Moreover, in contrast to $\kappa_0$ and the $\sigma_{m j}$
whose expressions are invariant by AB-exchange,  the coefficients $\beta_j$ or $\sigma_j$ may be non-symmetric in general, which would ensure that even if $g_A$ = $g_B$, the conductivity can become  dependent on the direction of the flux.


Let us define the fluxes $J_{AB}$ and $J_{BA}$ for the 
positive temperature gradient ($T_A=T_{h} > T_{c} =T_B$) and the reversed one, as schematized in Fig.~\ref{fig:two-particle-model}. From the expression of  $J$, we have
\begin{eqnarray} \nonumber
J_{AB}\equiv J_{A\to B} &=&  \big[ \kappa_0 + \epsilon 
\underbrace{\kappa_1(T_{h},T_{c})}_{\displaystyle \kappa_1^{AB}} \big](T_h-T_c),\\ \nonumber
J_{BA}\equiv  J_{B\to A} &=&  \big[ \kappa_0 + \epsilon 
\underbrace{\kappa_1(T_{c},T_{h})}_{\displaystyle \kappa_1^{BA}} \big](T_c-T_h). \\
\end{eqnarray}
Rectification emerges when $\kappa_1^{AB} \neq \kappa_1^{BA}$, molded by the functions $g_j(z)$, associated to nonlinear forces $f_j(z)$, which introduce the dependence of the conductivity on the bath temperatures.

In what follows, to quantify the diode effect,  
we use the ratio  
\begin{eqnarray} \label{eq:xi}
\chi \equiv\frac{ \big| |J_{AB}| - |J_{BA}| \big| }{\big( |J_{AB}| + |J_{BA}|\big)/2}  =  \epsilon  \frac{  |\kappa_1^{AB} - \kappa_1^{BA}|}{\kappa_0} + O(\epsilon^2) \,. 
\end{eqnarray}
This quantity coincides with the rectification factor~\cite{casati-diode} at first order in $\epsilon$ and it is twice the diodicity~\cite{alexander}. 
Notice that, the departure from the linear regime, signaled by 
$\epsilon\neq0$, together with asymmetry, is required to allow the diode effect ($\chi \neq 0$) at first order in $\epsilon$.  
 However, the rectification $\chi$ is small, of order $\epsilon$.

In the following sections, we will discuss the behavior of $\chi$ in some particular cases, in order to reduce  the number of parameters.

\subsection{Symmetric chain} 

Let us address the case 
where  $k_A=k_B$, $m_A=m_B$ and 
$\gamma_A=\gamma_B$, hence,  the asymmetry required for rectification must reside in the nonlinear on-site forces.
In this simple case, the coefficients obtained in Appendix~\ref{app:heat} 
reduce to
    \begin{eqnarray}
    \sigma_m &=& \frac{[ ( k+k_I)\bar{m}+1]k_I^2}{k(k+2k_I)( k+k_I  +\bar{m} k_I^2)}  \,,
    \\ \sigma_A&=&\sigma_B=\sigma=\frac{1}{k+k_I +\bar{m} k_I^2}  \,,\\
   \beta_A&=&\beta_B =  \sigma/2. 
    \end{eqnarray} 
Then, 
\begin{eqnarray} \label{eq:xik}
\chi&=& \epsilon\frac{\big| g_A(T_+)-g_A(T_-)
 +g_B(T_-)-g_B(T_+)
\big|}{2[ k+k_I  + \bar{m} k_I^2]}\,,
\end{eqnarray}
where $T_\pm
=\sigma_m T_m +\sigma T_{\substack{ h\\[-2pt]c}}=
(\sigma_m+\sigma)T_m \pm \sigma (T_h-T_c)/2$.

\begin{figure}[h!]
\centering 
\includegraphics[width = 0.45\textwidth]{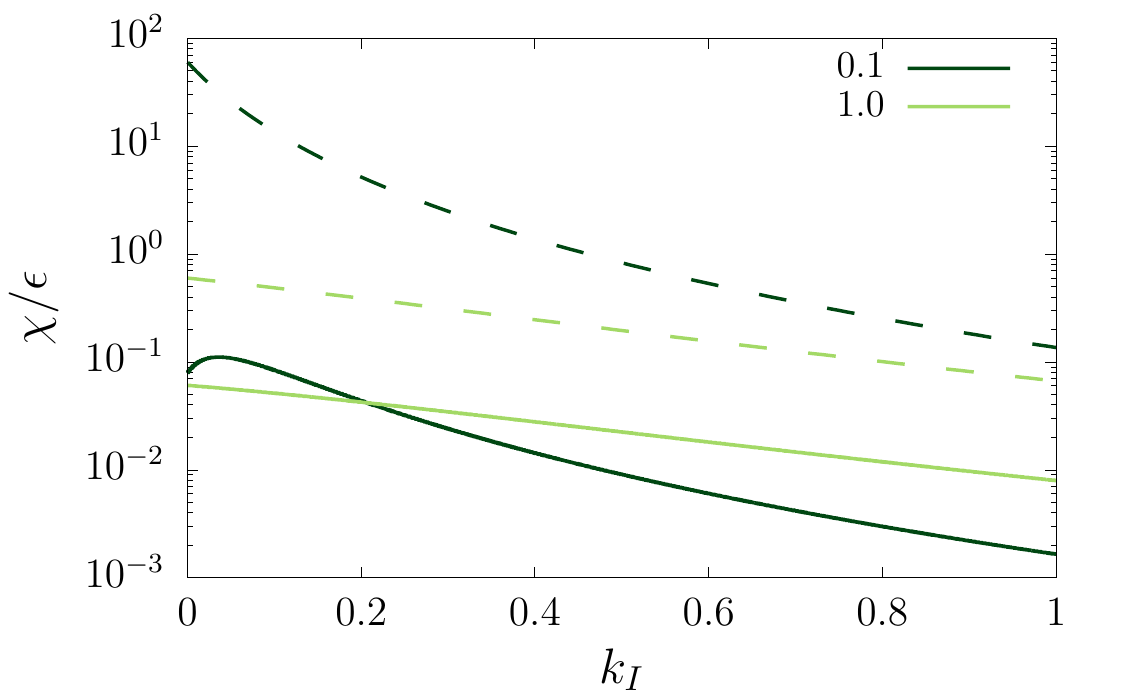}
\caption{Scaled rectification factor as a function of the interfacial stiffness $k_I$, for the nonlinear on-site potential $V_A^{nl}(z)=z^4/4$, (power-law, dashed lines), and 
$V_A^{nl}(z)=-\cos(z)$, (sinusoidal, solid lines), for $k=0.1$ (dark green) and 1 (light green), as indicated in the legend.  
In all cases $V_B^{nl}\equiv 0$,  $\bar{m}=1$, $T_m=1$, $\Delta T =0.4$. 
(Color online.)
}
\label{fig:xi1} 
\end{figure}

First, we notice that $\chi$ is finite in the 
limit $k_I \to 0$ and it tends to zero in the opposite limit $k_I\to \infty$. 
Examples are given in Fig.~\ref{fig:xi1} 
for two different potentials.

\begin{figure}[h!]
\centering 
\includegraphics[width = 0.45\textwidth]{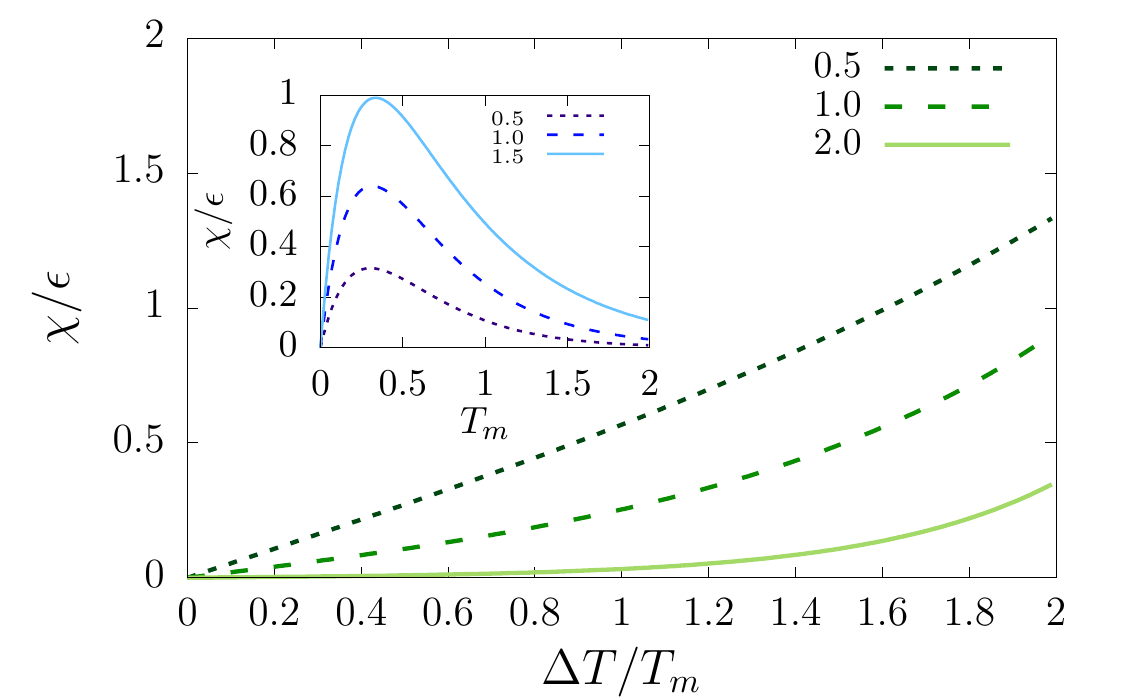}
\caption{Scaled rectification factor as a function of the relative temperature difference $\Delta T/T_m$, for the nonlinear on-site potential 
$V_A^{nl}(z)=-\cos(z)$,   
 $V_B^{nl}\equiv 0$, $k=k_I =0.1$, 
for different values of $T_m$, indicated in the legend. 
The inset displays the scaled rectification 
vs. $T_m$ for different values of $\Delta T$ indicated in the legend, showing that there is an optimal $T_m$, due to the loss of nonlinearity in the extremes of low and high temperatures for the chosen potential. (Color online.)
}
\label{fig:T1} 
\end{figure}

Rectification enhancement can be achieved by augmenting the temperature difference, fixing the average, since $\Delta g_j \equiv  g_j(T^+)-g_j(T^-) 
\approx  g_j'([\sigma_m+\sigma]T_m)\, \Delta T +  {\cal O}([\Delta T]^3)$. 
This effect is illustrated inf Fig.~\ref{fig:T1}. 
As a matter of fact, the increase of the rectification factor with the temperature difference has been observed in diverse models~\cite{efficiency2,efficiency3,bastida}.

 The mass and inverse square damping contribute trough $\bar{m}=m/\gamma^2$ to spoil rectification if $k_I>0$. This suggests that the overdamped regime  would perform  rectification better.

We can also understand 
how the  preferential direction in which the conductivity is larger, for given bath temperatures, depends on the type of nonlinear forces. 
For instance,  let us consider $g_B(z)=0$.  
If $g_A(z)$ is monotonically decreasing, 
like in the power-law case of Table I, then $\Delta g_A < 0$, for $T_A>T_B$, indicating that the preferential direction is from B to A (in general from smaller to larger nonlinear force). However,  if the potential is sinusoidal,  $g_A(z)$ is an increasing function (Table I), then the preferential direction is inverted with respect to the previous case (i.e., it is from A to B), as observed for asymmetric FK chains~\cite{casati-diode}).

Let us take a closer look to the conductivity in the limit 
$k_I \to 0$, for some concrete potentials $V_A^{nl}$, 
while $V_B^{nl}=0$. 
Recall that the conductivity scales with $k_I^2$, then the fluxes vanish in the limit $k_I\to 0$, 
however $\chi$ can be large for finite but very small $k_I$. 
In that limit, we have $\sigma_m=0$, $\sigma=1/k $, 
hence the scaled mass $\bar{m}$ does not play a role in the rectification. 

For the power-law on-site potential  $V_A^{nl}(z)= z^{2n}/(2n)$, $g_A(z)=-(2n-1)!!  z^{n-1}$, with  $n>1$, in the limit $k_I \to 0$,
Eq.~(\ref{eq:xik}) becomes
\begin{eqnarray} \label{eq:limxi0a}
\chi=   \epsilon  \frac{ (2n-1)!! }{ 2 k }\left[ 
\left(\frac{T_h}{k}\right)^{n-1} - \left(\frac{T_c}{k}\right)^{n-1} \right] \,.
\end{eqnarray}

Equation~(\ref{eq:limxi0a}) predicts, for instance, that the ratio $\chi$  grows with the temperature difference $\Delta T= T_h-T_c$,  with positive concavity for $n>2$, nearly linear for small $\Delta T$. 
 These effects persist for finite $k_I$ as shown in Fig.~\ref{fig:T1}.
 
For the sinusoidal on-site potential  $V_A^{nl}(z)=- \cos(K z)/K $  
(as in the FK model),   $g_A(z)=-   K  { e}^{-K^2 z/2}$ (see Table I). In this case the dependence $\chi$ vs. $k_I$ can be nonmonotonic, with a finite optimal value, as shown in Fig.~\ref{fig:xi1}.  In the limit $k_I\to 0$ we obtain
\begin{eqnarray}
\chi= \epsilon \frac{K}{2k} 
\left( e^{-K^2T_c/(2k)} - e^{-K^2T_h/(2k)}\right) \,.
\end{eqnarray}
 
The dependence on $\Delta T$ (for fixed mean temperature $T_m$) is also an increasing convex function. 
This behavior, which also holds for finite $k_I$, as exemplified in Fig.~\ref{fig:T1}, is qualitatively similar to that  reported from simulations of diode models~\cite{efficiency2,efficiency3,bastida}.
For the power-law potential the nonlinear correction is  weak but in the same direction.  
 
\subsection{Small $k_I$ limit}

\begin{figure}[h!]
\centering 
\includegraphics[width = 0.45\textwidth]{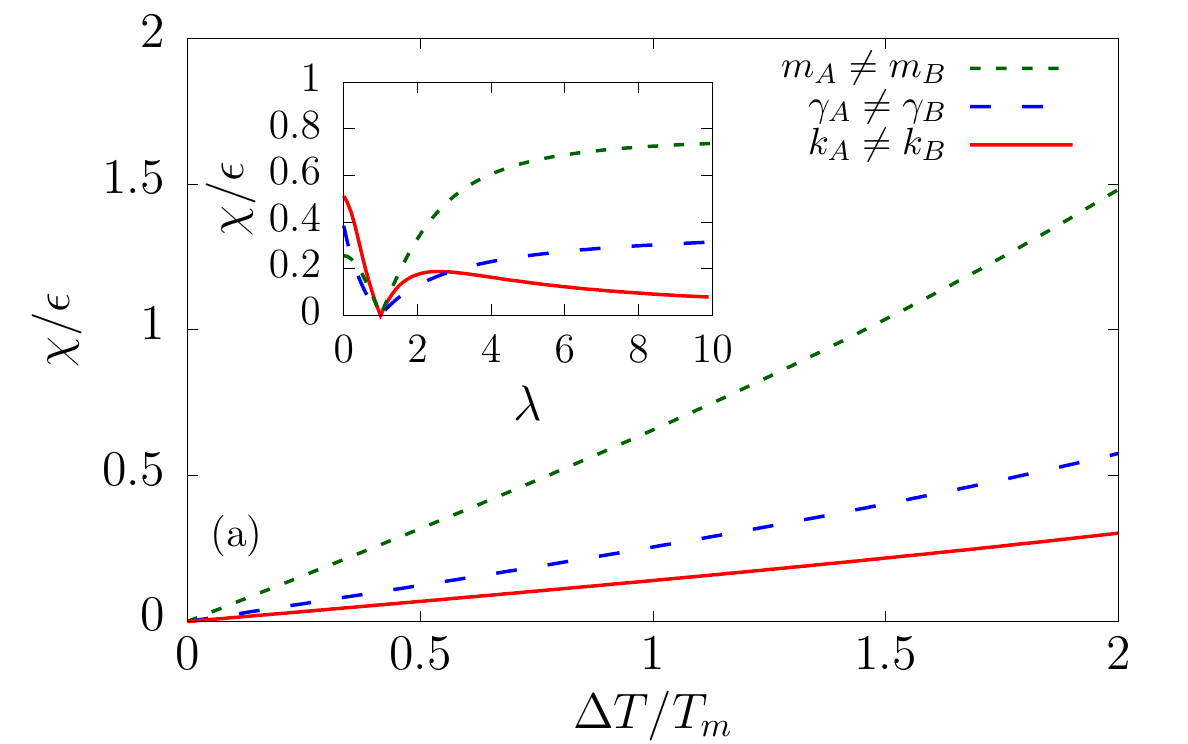}
\includegraphics[width = 0.45\textwidth]{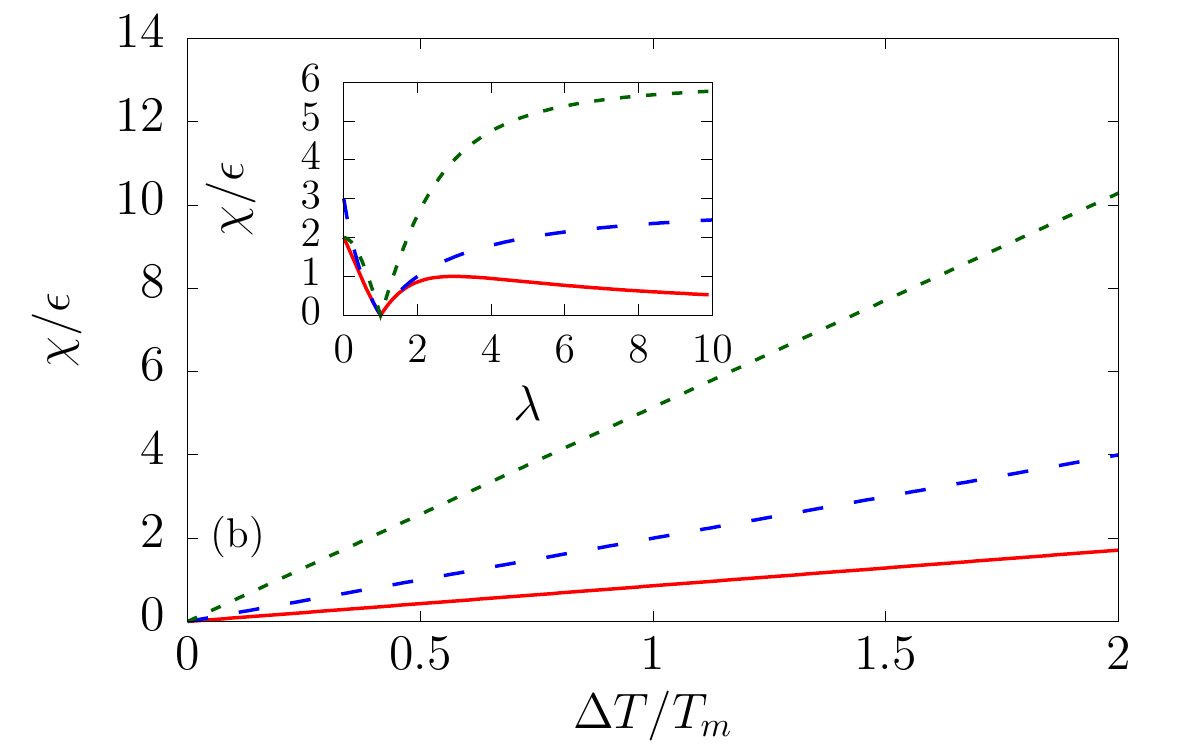}
\caption{Scaled rectification factor vs. the relative temperature difference $\Delta T/T_m$, with $T_m=1$, for  nonlinear on-site potential  
$V_A^{nl}(z)=V_B^{nl}(z)=$ (a) $-\cos(z)$ and 
(b)  $z^4/4$  for:
  $m_A=5$, $k_A=k_B=m_B=\gamma_A=\gamma_B=1$ (green),
  $\gamma_A=5$, $\gamma_B=k_A=k_B=m_A=m_B=1$ (blue) $k_A=5$, $k_B=m_A=m_B=\gamma_A=\gamma_B=1$ (red).
  The insets show the dependence on the asymmetry factor $\lambda$ that for each parameter $p$ gives  
  $p_A =\lambda p_B=\lambda$. 
(Color online.)
  }
\label{fig:all} 
\end{figure}

In the previous section, we have seen that the limit of small $k_I$ is relevant, while it allows to simplify the analytical expressions significantly. 
Then, in this limit, we will analyze the effect of introducing the asymmetry alternatively in the
stiffness ($k_A\neq k_B$),  mass ($m_A\neq m_B$) or damping coefficient ($\gamma_A\neq \gamma_B$). 
As shown  in Appendix~\ref{app:smallkI}, 
in this limit, 
\begin{eqnarray}
\kappa_1 & \simeq & \kappa_0\Big[\beta_A\, g_A(\sigma_A T_A)   +  
   \beta_B \,  g_B( \sigma_B T_B)  \Big] \,,
\end{eqnarray}
where the explicit expressions for the coefficients are given in the Appendix~\ref{app:smallkI} for each asymmetry. 
The results for the rectification factor $\chi$ are illustrated in Fig.~\ref{fig:all}. 

{\bf Note:} We observe that any of these asymmetries can produce rectification. In particular, notice that, even when the chain is homogeneous, distinct thermostats (characterized by different friction coefficients) can also produce a diode effect. 
 
\section{Final remarks}
\label{sec:final}

We have presented analytical results starting from the microscopic classical dynamics of  
a two-particle system with nonlinear forces. 
Due to the nonlinearity of the equations, we tackled the solution from a perturbative approach valid for small nonlinear intensity $\epsilon$. 
It is noticeable  that despite the simplicity of the system,   the conductivity
$\kappa = \kappa_0 + \epsilon \kappa_1$
has an intricate dependence on the system parameters. Therefore, it might be hard to make a portrait of this complexity only through molecular dynamics simulations,  
making worthy the present effort of obtaining analytical results from first principles.  

Some previously known results can be revisited from this perspective. 
Particularly,  
one can see how the temperature-dependence of the conductivity emerges from the nonlinearity of the forces,  through the functions $g_j(z)$. 
The requirements  of broken symmetry and of nonlinearity explicitly  appear. 
The results also allow shedding light on effects observed in 
chains, e.g., the scaling of the conductivity with the 
interfacial stiffness $k_I$,  the dependence of the rectification factor on $k_I$ and 
 on the temperature difference. 
How nonlinearities determine the preferential direction has also been foreseen. The role of different asymmetries (in the mass, stiffness, on-site potential and even damping coefficient) was also shown. 

It is interesting to note that, from Appendix~\ref{app:heat}, it is possible to obtain that the nonlinearity yields a temperature-dependent power spectrum (anharmonic phonons), which can be seen as the correction to the harmonic theory responsible for phonon scattering~\cite{hanggi}. 
The relationship between temperature and the overlapping phonon bands has already been analytically studied for FK asymmetric chains \cite{casati-diode} and chains with dissimilar  anharmonic segments (FK and Fermi-Pasta-Ulan-Tsingou) ~\cite{casati-diode,BaowenLi2005}. 

Our results are valid when the effect of the nonlinear forces can be treated as a perturbation to the predominantly linear solutions. Consequently, the predicted diode effect is very small. 
However, the results allow for a clear glance regarding the mechanisms behind rectification and the role of diverse asymmetries and nonlinearities. 
 
Possible extensions include baths of different nature (correlated or non-Gaussian) and nonlinear interfacial interactions.

{\bf Acknowledgments:}  
We are grateful to Alexandre Almeida for the fruitful discussions. CA acknowledges Brazilian agency CNPq (process 311435/2020-3) for partial financial support.  CAPES (finance code 001) is also acknowledged.

\bibliographystyle{plain}

\onecolumngrid

\appendix
\setcounter{figure}{0}
\setcounter{equation}{0}
\renewcommand{\theequation}{A\arabic{equation}}
\renewcommand{\thefigure}{A\arabic{figure}}

\setcounter{subsection}{0}
\renewcommand\thesubsection{A\thesection\arabic{subsection}}

\section*{Appendix}

\subsection{Perturbative solution}
\label{app:xy}
 
\subsubsection{Zeroth order solution}

Eqs.~(\ref{eq:0x}) and (\ref{eq:0y})   in matrix form are
\begin{eqnarray}
\begin{pmatrix}
a(\omega) & - k_I \\
- k_I & b(\omega)
\end{pmatrix}
\begin{pmatrix}
\tilde{x}_0(\omega) \\
\tilde{y}_0 (\omega)
\end{pmatrix} = \begin{pmatrix}
\tilde{\eta}_A(\omega) \\
\tilde{\eta}_B (\omega)
\end{pmatrix},
\end{eqnarray}
whose solution, by matrix inversion, is 
\begin{eqnarray}
\begin{pmatrix}
\tilde{x}_0(\omega) \\
\tilde{y}_0 (\omega)
\end{pmatrix} =\frac{1}{a(\omega)b(\omega)-k_I^2} 
\begin{pmatrix}
b(\omega) & k_I\\
k_I & a(\omega)
\end{pmatrix}
\begin{pmatrix}
\tilde{\eta}_A(\omega) \\
\tilde{\eta}_B (\omega)
\end{pmatrix}
=
\begin{pmatrix}
 M_{11}(\omega) &  M_{12}(\omega) \\
 M_{21}(\omega) &  M_{22}(\omega)
\end{pmatrix}
\begin{pmatrix}
\tilde{\eta}_A(\omega) \\
\tilde{\eta}_B (\omega)
\end{pmatrix} \, ,
\end{eqnarray}
corresponding  to Eq.~(\ref{eq:xyw0}). The Fourier transforms $\tilde{\eta}_A (\omega)$ and $\tilde{\eta}_B(\omega)$ are also Gaussian distributed with
\begin{eqnarray}
\langle \tilde{\eta}_A (\omega) \rangle &=& \langle \tilde{\eta}_B(\omega) \rangle = 0 \,, \\
\langle \tilde{\eta}_A(\omega) \tilde{\eta}_A (\omega') &=& 2 \gamma T_A \big[ 2 \pi \delta(\omega+\omega') \big] \,, \\
\langle \tilde{\eta}_B(\omega) \tilde{\eta}_B (\omega') &=& 2 \gamma T_B \big[ 2 \pi \delta(\omega+\omega') \big] \, .
\end{eqnarray}

Using the solutions and the correlations of the noises, we can calculate the coordinate correlations in Fourier space, namely
\begin{eqnarray}
	\langle \tilde{x}_0 (\omega) \tilde x_0 (\omega') \rangle &=& 2 \gamma \left\{   M_{11}(\omega)  M_{11}(\omega') T_A +  M_{12}(\omega)  M_{12}(\omega') T_B  \right\} 2 \pi \delta(\omega+\omega') \,,  \\
	\langle \tilde{y}_0 (\omega) \tilde y_0 (\omega')  \rangle  &=&  2 \gamma \left\{   M_{21}(\omega)  M_{21}(\omega') T_A +  M_{22}(\omega)  M_{22}(\omega') T_B  \right\} 2 \pi \delta(\omega+\omega') \,, \\ \label{eq:xy}
	\langle \tilde{x}_0 (\omega) \tilde y_0 (\omega') \rangle &=&  2 \gamma \left\{   M_{11}(\omega)  M_{21}(\omega') T_A +  M_{12}(\omega)  M_{22}(\omega') T_B  \right\} 2 \pi \delta(\omega+\omega') \, .
\end{eqnarray}

\subsubsection{First order}

The Fourier transformed Eqs.~(\ref{eq:x1}) and (\ref{eq:y1}) 
in matrix form are 
\begin{eqnarray}
\begin{pmatrix}
a(\omega) & - k_I \\
- k_I & b(\omega)
\end{pmatrix}
\begin{pmatrix}
\tilde{x}_1(\omega) \\
\tilde{y}_1 (\omega)
\end{pmatrix} = \begin{pmatrix}
\mathcal{F} \left\{  f_A({x}_0) \right\} (\omega) \\
\mathcal{F} \left\{  f_B({y}_0) \right\} (\omega)
\end{pmatrix}
\end{eqnarray}
and their respective solutions are
\begin{eqnarray}
\begin{pmatrix}
\tilde{x}_1(\omega) \\
\tilde{y}_1 (\omega)
\end{pmatrix}
=
\begin{pmatrix}
 M_{11}(\omega) &  M_{12}(\omega) \\
 M_{21}(\omega) &  M_{22}(\omega)
\end{pmatrix}
\begin{pmatrix}
\mathcal{F} \left\{  f_A({x}_0) \right\} (\omega) \\
\mathcal{F} \left\{  f_B({y}_0) \right\} (\omega)
\end{pmatrix} \, ,
\end{eqnarray}
giving Eq.~(\ref{eq:xyw1}).
Recalling that $f_i(z)=\sum_n c_{i,n} z^n$, for $i=A,B$, then 
\begin{eqnarray}
{\cal F} \{ f_i(z) \} &=& \sum_n c_{i,n} {\cal F}\{ z^n\}(\omega) \, ,
\end{eqnarray}
leading to
\begin{eqnarray} \nonumber
\tilde{x}_1(\omega) &=&  M_{11}(\omega) {\cal F}\{f_A(x_0)\}(\omega) +  M_{12} (\omega) {\cal F}\{f_B(y_0)\}(\omega) \\
 &=&  M_{11}(\omega) \sum_n c_{A,n} {\cal F}\{x_0^n\}(\omega) +  M_{12} (\omega) \sum_n c_{B,n} {\cal F}\{ y_0^n\}(\omega) \,, \\ \nonumber
\tilde{y}_1(\omega) &=&  M_{21}(\omega) {\cal F}\{ f_A(x_0)\}(\omega) +  M_{22} (\omega) {\cal F} \{ f_B(y_0) \}(\omega)   \\
  &=&  M_{21}(\omega) \sum_n c_{A,n} {\cal F}\{x_0^j\}(\omega) +  M_{22} (\omega) \sum_n c_{B,n} {\cal F}\{ y_0^j \}(\omega) \, .
\end{eqnarray}

Moreover, we will use below that the Fourier transform of an integer power of a function $z(t)$ is
\begin{eqnarray} \label{eq:fourierpower}
		{\cal F}\{  z^n \} (\omega) &=& \int \prod_{j=1}^{n} \frac{d\omega_j}{2 \pi} \delta\biggl(\omega - \sum_{j=1}^n \omega_j\biggr) \prod_{j=1}^{n} \tilde{z} (\omega_j) \, .
	\end{eqnarray}

\subsection{Heat flow}
\label{app:heat}

\subsubsection{Zeroth order}

From Eq.\,(\ref{eq:Jdef}), using the first definition and  writing the average in Fourier space
\begin{eqnarray}
	J_x = k_I \langle (x_0 - y_0) \dot{x}_0 \rangle  = k_I \int \frac{d \omega d \omega'}{(2 \pi)^2} e^{i t (\omega + \omega')}   \langle (\tilde x_0(\omega) - \tilde y_0 (\omega)) (i \omega' \tilde{x}_0(\omega')) \rangle \, ,
\end{eqnarray}  
the first term is null 
since $\langle \tilde{x}_0 (\omega) \tilde{x}_0 (-\omega) \rangle$ is an even function, when we multiply times $i \omega$ from the time derivative, the function becomes odd and as we integrate for all $\omega$, the value becomes null.  Then, using Eq.~(\ref{eq:xy}),
\begin{eqnarray} \nonumber
	J_0 &=& J_x  = - k_I \int \frac{d \omega}{2 \pi} e^{it(\omega+\omega')} (i \omega') \langle \tilde{y}_0(\omega) \tilde{x}_0(\omega') \rangle  \\
	&=& k_I \int \frac{d \omega}{2 \pi}  2 \gamma (i \omega)   \left\{   \frac{k_I \, b(-\omega) T_A + k_I \, a(\omega) T_B}{\big(a(\omega)b(\omega) - k_I^2\big)\big(a(-\omega)b(-\omega) - k_I^2\big)}  \right\} \, ,
\end{eqnarray}
which, if we perform a change of variables ($\omega \to - \omega$) in the second term gives
\begin{eqnarray}  
	J_0	&=&k_I \int \frac{d \omega}{2 \pi}  2 \gamma (i \omega)   \left\{   \frac{k_I \, b(-\omega) T_A - k_I \, a(-\omega) T_B}{\big(a(\omega)b(\omega) - k_I^2\big)\big(a(-\omega)b(-\omega) - k_I^2\big)}  \right\} \, ,
\end{eqnarray}
where only the odd parcel of the numerator (since the denominator is always even) will yield a nonnull result after the integration. We expand the expression to
\begin{eqnarray} 
	J_0 &=&  \int \frac{d \omega}{2 \pi}    \left\{   \frac{2 \gamma^2 k_I^2 \omega^2 }{\big(a(\omega)b(\omega) - k_I^2\big)\big(a(-\omega)b(-\omega) - k_I^2\big)}  \right\}  \big(T_A - T_B\big)  \equiv \kappa_0\, \big(T_A - T_B\big)\, ,
\end{eqnarray}
corresponding to Eqs. (\ref{eq:Jx}) and   (\ref{eq:kappa0}). 

\subsubsection{First order}

The correlations required to compute $J_1$ are
\begin{eqnarray} \nonumber
	 \langle y_1 \dot{x}_0 \rangle  &=& \int \frac{d \omega d \omega'}{(2 \pi)^2} e^{it(\omega+\omega')} (i \omega') \langle \tilde{y}_1(\omega) \tilde{x}_0(\omega') \rangle  \\
	&=& \int \frac{d \omega d \omega'}{(2 \pi)^2} e^{it(\omega+\omega')} (i \omega') \sum_ n\left[ c_{A,n}  M_{21}(\omega) \left\langle {\cal F}\{ x_0^n \}(\omega) \tilde{x}_0 (\omega') \right\rangle + c_{B,n}  M_{22}(\omega) \left\langle {\cal F}  \{ y_0^ n \} (\omega) \tilde{x}_0 (\omega') \right\rangle \right] \, ,
\end{eqnarray}
and
\begin{eqnarray} \nonumber
	 \langle y_0 \dot{x}_1 \rangle  &=& \int \frac{d \omega d \omega'}{(2 \pi)^2} e^{it(\omega+\omega')} (i \omega') \langle \tilde{y}_0(\omega) \tilde{x}_1(\omega') \rangle  \\
	&=& \int \frac{d \omega d \omega'}{(2 \pi)^2} e^{it(\omega+\omega')} (i \omega') \sum_n \left[ c_{A,n}  M_{11}(\omega') \left\langle {\cal F}\{ x_0^ n \} (\omega') \tilde{y}_0 (\omega) \right\rangle + c_{B,n}  M_{12}(\omega') \left\langle {\cal F}\{ y_0^ n \} (\omega') \tilde{y}_0 (\omega) \right\rangle \right] \, .
\end{eqnarray}

First, we evaluate the following correlations between the zeroth-order terms,
\begin{eqnarray} \nonumber
\int \frac{d\omega_1 d \omega_2}{(2 \pi)^2} \langle \tilde{x}_0(\omega_1) \tilde{x}_0 (\omega_2) \rangle &=& 2 \gamma \int \frac{d\omega}{2\pi} \big\{  M_{11}(\omega) M_{11}(-\omega) T_A +  M_{12}(\omega) M_{12}(-\omega) T_B  \big\} \\ \label{eq:sigmaA}
&=& \sigma_{mA}T_m +  \sigma_A T_A \, ,
\end{eqnarray}
\begin{eqnarray} \nonumber
\int \frac{d\omega_1 d \omega_2}{(2 \pi)^2} \langle \tilde{y}_0(\omega_1) \tilde{y}_0 (\omega_2) \rangle &=& 2 \gamma \int \frac{d\omega}{2\pi} \big\{  M_{21}(\omega) M_{21}(-\omega) T_A +  M_{22}(\omega) M_{22}(-\omega) T_B  \big\}\\  \label{eq:sigmaB}
&=& \sigma_{mB} T_m+  \sigma_B T_B \, .
\end{eqnarray}
These correlations, together with  Eq.~(\ref{eq:fourierpower}), will be used to evaluate the    correlations required to compute $J_1$, namely, 
\begin{eqnarray} \nonumber
\langle {\cal F}(x_0^n)(\omega) \tilde x _0 (\omega') \rangle  &=& \int \biggl[ \prod_{j=1}^{n} \frac{d\omega_j}{2 \pi} \biggr] \delta \biggl(\omega - \sum_{j=1}^n \omega_j\biggr)  \left\langle \biggl[  \prod_{j=1}^{n} \tilde{x}_0 (\omega_j) \biggr] \tilde{x}_0(\omega') \right\rangle \\ \nonumber
&=& n!! \int \biggl[ \prod_{j=1}^{n} \frac{d\omega_j}{2 \pi} \biggr] \delta\biggl(\omega - \sum_{j=1}^n \omega_j\biggr)  \langle \tilde x_0 (\omega_j) \tilde x_0 (\omega')  \rangle \prod_{j \text{ odd} }^{n-1}  \langle \tilde x_0 (\omega_j) \tilde x_0 (\omega_{j+1})  \rangle  \\ \nonumber
&=& n!! \, 2 \pi \delta(\omega+\omega') \int \frac{d \omega_n}{2 \pi}  \langle \tilde x_0 (\omega_n) \tilde x_0 (\omega')  \rangle \prod_{j \text{ odd} }^{n-1} \int \frac{d\omega_j d\omega_{j+1}}{(2 \pi)^2} \left\langle \tilde x_0 (\omega_j) \tilde x_0 (\omega_{j+1}) \right\rangle \\
&=& n!! \, 2 \pi \delta(\omega + \omega') \int \frac{d \omega_n}{2 \pi} \left\langle \tilde x_0 (\omega_n) \tilde x_0 (\omega') \right\rangle \Big[ \sigma_{mA} T_m+  \sigma_A T_A \Big]^{\frac{n-1}{2}} \,.
\end{eqnarray}

\begin{eqnarray}\nonumber
\langle {\cal F}(x_0^n)(\omega) \tilde y _0 (\omega') \rangle  &=& \int \biggl[ \prod_{j=1}^{n} \frac{d\omega_j}{2 \pi} \biggr] \delta\biggl(\omega - \sum_{j=1}^n \omega_j\biggr)  \left\langle \biggl[  \prod_{j=1}^{n} \tilde{x}_0 (\omega_j) \biggr] \tilde{y}_0(\omega') \right\rangle \\ \nonumber
&=& n!! \int\biggl[ \prod_{j=1}^{n} \frac{d\omega_j}{2 \pi} \biggr] \delta\biggl(\omega - \sum_{j=1}^n \omega_j\biggr) \left\langle \tilde x_0 (\omega_n) \tilde y_0 (\omega') \right\rangle \prod_{j \text{ odd} }^{n-1} \left\langle \tilde x_0 (\omega_j) \tilde x_0 (\omega_{j+1}) \right\rangle  \\ \nonumber
&=& n!! \, 2 \pi \delta(\omega+\omega') \int \frac{d \omega_n}{2 \pi} \left\langle \tilde x_0 (\omega_n) \tilde y_0 (\omega') \right\rangle \prod_{j \text{ odd} }^{n-1} \int \frac{d\omega_j d\omega_{j+1}}{(2 \pi)^2} \left\langle \tilde x_0 (\omega_j) \tilde x_0 (\omega_{j+1}) \right\rangle  \\
&=& n!! \, 2 \pi \delta(\omega + \omega') \int \frac{d \omega_n}{2 \pi} \left\langle \tilde x_0 (\omega_n) \tilde y_0 (\omega') \right\rangle \Big[ \sigma_{mA}T_m +  \sigma_A T_A \Big]^{\frac{n-1}{2}} \,.
\end{eqnarray}

\begin{eqnarray}\nonumber
\langle {\cal F}(y_0^n)(\omega) \tilde x _0 (\omega') \rangle  &=&\int \biggl[ \prod_{j=1}^{n} \frac{d\omega_j}{2 \pi} \biggr] \delta\biggl(\omega - \sum_{j=1}^n \omega_j\biggr)  \left\langle\biggl[  \prod_{j=1}^{n} \tilde{y}_0 (\omega_j) \biggr] \tilde{x}_0(\omega') \right\rangle \\ \nonumber
&=& n!! \int \biggl[ \prod_{j=1}^{n} \frac{d\omega_j}{2 \pi} \biggr] \delta\biggl(\omega - \sum_{j=1}^n \omega_j\biggr) \left\langle \tilde y_0 (\omega_n) \tilde x_0 (\omega') \right\rangle \prod_{j \text{ odd} }^{n-1} \left\langle \tilde y_0 (\omega_j) \tilde y_0 (\omega_{j+1}) \right\rangle  \\ \nonumber
&=& n!! \, 2 \pi \delta(\omega+\omega') \int \frac{d \omega_n}{2 \pi} \left\langle \tilde y_0 (\omega_n) \tilde x_0 (\omega') \right\rangle \prod_{j \text{ odd} }^{n-1} \int \frac{d\omega_j d\omega_{j+1}}{(2 \pi)^2} \left\langle \tilde y_0 (\omega_j) \tilde y_0 (\omega_{j+1}) \right\rangle  \\
&=& n!! \, 2 \pi \delta(\omega + \omega') \int \frac{d \omega_n}{2 \pi} \left\langle \tilde y_0 (\omega_n) \tilde x_0 (\omega') \right\rangle \Big[ \sigma_{mB} T_m+  \sigma_B T_B \Big]^{\frac{n-1}{2}} \,.
\end{eqnarray}

\begin{eqnarray}\nonumber
\langle {\cal F}(y_0^n)(\omega) \tilde y _0 (\omega') \rangle  &=& \int \biggl[ \prod_{j=1}^{n} \frac{d\omega_j}{2 \pi} \biggr] \delta\biggl(\omega - \sum_{j=1}^n \omega_j\biggr)  \left\langle \biggl[  \prod_{j=1}^{n} \tilde{y}_0 (\omega_j) \biggr] \tilde{y}_0(\omega') \right\rangle \\ \nonumber
&=& n!! \int \biggl[ \prod_{j=1}^{n} \frac{d\omega_j}{2 \pi} \biggr] \delta\biggl(\omega - \sum_{j=1}^n \omega_j\biggr) \left\langle \tilde y_0 (\omega_n) \tilde y_0 (\omega') \right\rangle \prod_{j \text{ odd} }^{n-1} \left\langle \tilde y_0 (\omega_j) \tilde y_0 (\omega_{j+1}) \right\rangle  \\ \nonumber
&=& n!! \, 2 \pi \delta(\omega+\omega') \int \frac{d \omega_n}{2 \pi} \left\langle \tilde y_0 (\omega_n) \tilde y_0 (\omega') \right\rangle \prod_{j \text{ odd} }^{n-1} \int \frac{d\omega_j d\omega_{j+1}}{(2 \pi)^2} \left\langle \tilde y_0 (\omega_j) \tilde y_0 (\omega_{j+1}) \right\rangle  \\
&=& n!! \, 2 \pi \delta(\omega + \omega') \int \frac{d \omega_n}{2 \pi} \left\langle \tilde y_0 (\omega_n) \tilde y_0 (\omega') \right\rangle \Big[ \sigma_{mB} T_m+  \sigma_B T_B \Big]^{\frac{n-1}{2}} \,.
\end{eqnarray}

Now we write Eq.\,(\ref{eq:J01}) as
\begin{eqnarray}  
	 && -k_I [ \langle y_1 \dot{x}_0 \rangle +  \langle y_0 \dot{x}_1 \rangle ] = \\ \nonumber
	 &&= - k_I \int \frac{d \omega d \omega'}{(2 \pi)^2} e^{it(\omega+\omega')} (i \omega') \sum_ n \Big\{  c_{A,n} \big[   M_{21}(\omega)  \langle {\cal F}\{ x_0^ n \}(\omega) \tilde{x}_0 (\omega')  \rangle + M_{11}(\omega')  \langle {\cal F}\{ x_0^ n \} (\omega') \tilde{y}_0 (\omega)  \rangle \big] + \\ \nonumber
	 && + c_{B,n} \big[  M_{22}(\omega)  \langle {\cal F}  \{ y_0^ n \} (\omega) \tilde{x}_0 (\omega')  \rangle  +   M_{12}(\omega')  \langle {\cal F}\{ y_0^ n \} (\omega') \tilde{y}_0 (\omega)  \rangle \big] \Big\} \, ,
\end{eqnarray}
grouping all terms proportional to $c_{A,n}$
\begin{eqnarray}\nonumber
&& - k_I \int \frac{d\omega d\omega'}{2\pi^2} e^{it(\omega+\omega')} (i \omega') \big[   M_{21}(\omega) \left\langle {\cal F}\{ x_0^n \}(\omega) \tilde{x}_0 (\omega') \right\rangle + M_{11}(\omega') \left\langle {\cal F}\{ x_0^n \} (\omega') \tilde{y}_0 (\omega) \right\rangle \big] =  \\ \nonumber
&=& - k_I  n!! \Big[ \sigma_{mA}T_m +  \sigma_A T_A \Big]^{\frac{n-1}{2}}  \int \frac{d \omega d \omega' d \omega_n}{(2 \pi)^2}  e^{it(\omega+\omega')} (i \omega') \delta(\omega + \omega') \big[   M_{21}(\omega) \left\langle \tilde x_0 (\omega_n) \tilde x_0 (\omega') \right\rangle +  M_{11}(\omega') \left\langle \tilde x_0 (\omega_n) \tilde y_0 (\omega) \right\rangle  \big]  \\ \nonumber
&=& - k_I  n!! \Big[ \sigma_{mA} T_m+  \sigma_A T_A \Big]^{\frac{n-1}{2}}  \int \frac{d \omega' d \omega_n}{(2 \pi)^2} (i \omega')\big[   M_{21}(-\omega') \left\langle \tilde x_0 (\omega_n) \tilde x_0 (\omega') \right\rangle +  M_{11}(\omega') \left\langle \tilde x_0 (\omega_n) \tilde y_0 (-\omega') \right\rangle  \big] \\ \label{eq:betaA}
&=& n!! \,\kappa_0\,\beta_A  \Big[ \sigma_{mA} T_m+  \sigma_A T_A \Big]^{\frac{n-1}{2}}  (T_A - T_B),
\end{eqnarray}
and to $c_{B,n}$
\begin{eqnarray} \nonumber
&& - k_I  \int \frac{d\omega d\omega'}{2\pi^2} e^{it(\omega+\omega')} (i \omega') \big[   M_{22}(\omega) \left\langle {\cal F}\{ y_0^n \}(\omega) \tilde{x}_0 (\omega') \right\rangle + M_{12}(\omega') \left\langle {\cal F}\{ y_0^n \} (\omega') \tilde{y}_0 (\omega) \right\rangle \big] =  \\ \nonumber
&=& - k_I n!! \Big[\sigma_{mB} T_m+  \sigma_B T_B \Big]^{\frac{n-1}{2}}  \int \frac{d \omega d \omega' d \omega_n}{(2 \pi)^2}  e^{it(\omega+\omega')} (i \omega') \delta(\omega + \omega') \big[   M_{22}(\omega) \left\langle \tilde y_0 (\omega_n) \tilde x_0 (\omega') \right\rangle +  M_{12}(\omega') \left\langle \tilde y_0 (\omega_n) \tilde y_0 (\omega) \right\rangle  \big] \nonumber \\ \nonumber
&=& - k_I n!! \Big[ T_m\sigma_{mB} +  \sigma_B T_B \Big]^{\frac{n-1}{2}}  \int \frac{d \omega' d \omega_n}{(2 \pi)^2} (i \omega')\big[   M_{22}(-\omega') \left\langle \tilde y_0 (\omega_n) \tilde x_0 (\omega') \right\rangle +  M_{12}(\omega') \left\langle \tilde y_0 (\omega_n) \tilde y_0 (-\omega') \right\rangle  \big] \\  \label{eq:betaB}
&=& n!! \,\kappa_0\,\beta_B \Big[ \sigma_{mB}T_m +  \sigma_B T_B \Big]^{\frac{n-1}{2}}  (T_A - T_B) \, ,
\end{eqnarray}
finally yielding the first-order correction

\begin{eqnarray} \nonumber
J_1 &=&  \kappa_0\,\beta_A \sum_ n \left\{ c^A_ n  n!! \big[ \sigma_{mA}T_m +  \sigma_A T_A \big]^{\frac{ n-1}{2}}  \right\} (T_A - T_B) 
+  \kappa_0\,\beta_B\sum_ n \left\{ c^B_ n   n!! \big[\sigma_{mB}  T_m+  \sigma_B T_B \big]^{\frac{ n-1}{2}} \right\}  (T_A - T_B) \\
&\equiv&  \kappa_0\,\Bigl[ \beta_A\, g_A(\sigma_{mA} T_m+  \sigma_A T_A)  +  
 \beta_B \, g_B(\sigma_{mB} T_m+  \sigma_B T_B)  \Bigr] (T_A - T_B) \,, \label{eq:J1geral}
\end{eqnarray}
where we have defined  
\begin{eqnarray}
   g_j(z)  = \sum_ n   c_{j, n} \, n!!\, z^{\frac{ n-1}{2}} , 
\end{eqnarray}
for $j=A,B$.
Examples of the correspondence between $g_j$ and the force $f_j$ are given in Table I of the main text. 

Moreover, recall that $\kappa_0$ was defined in Eq.~(\ref{eq:J0}), and  the remaining coefficients in Eq.~(\ref{eq:J1geral}) were defined in Eqs.~(\ref{eq:sigmaA}), (\ref{eq:sigmaB}), (\ref{eq:betaA}) and (\ref{eq:betaB}). 
Their, explicit values for the particular case $m_A=m_B=m$, $\gamma_A=\gamma_B=\gamma$, additionally defining 
$\bar{m}=m/\gamma^2$,  are given below. 

\begin{eqnarray} 
 \kappa_0 &=&  \frac{2 k_I^2/\gamma }
{ \big(4k_I^2 + (k_A - k_B)^2\big) \bar{m} + 2( k_A+k_B+2k_I )}\,,  
\;\;\;\; \label{eq:kappa0explicit}
\end{eqnarray}

\begin{eqnarray}
\sigma_{mA}=\sigma_{mB} =	\sigma_m &=&  \frac{
 2 	k_I^2 \Big( \big( 2k_I + k_A+k_B \big)\bar{m} + 2  \Big)  }
	{
	\big( k_A k_B + k_I (k_A+k_B) \big) \Big( \big(4 k_I^2 + (k_A-k_B)^2\big) \bar{m} + 2\big(2 k_I + k_A+k_B \big)   \Big)  } \, , 
\end{eqnarray}
\begin{eqnarray}
	 \sigma_A &=& \frac{(k_A-k_B)
	 \big( (k_A- k_B)(k_B + k_I) - 2 k_I^2 \big)\bar{m}
	 + 2 \big( k_B (k_A + k_B) + k_I (k_A + 3 k_B) \big)  }
	 {\big(k_Ak_B  + k_I  (k_A+k_B) \big) 
	 \Big( \big(4 k_I^2 + (k_A-k_B)^2 \big) \bar{m} + 2\big(2 k_I +k_A+k_B \big)  \Big)} \, ,
\end{eqnarray}
\begin{eqnarray}
	 \sigma_B &=& \frac{(k_B-k_A)
	 \big( (k_B- k_A)(k_A + k_I) - 2 k_I^2 \big)\bar{m}
	 + 2 \big( k_A (k_A + k_B) + k_I (k_B + 3 k_A) \big)  }
	 {\big(k_Ak_B  + k_I  (k_A+k_B) \big) 
	 \Big( \big(4 k_I^2 + (k_A-k_B)^2 \big) \bar{m} + 2\big(2 k_I +k_A+k_B \big)  \Big)} \, 
\end{eqnarray}
\begin{eqnarray}
	\beta_A &=& \frac{2  \big(1 + \bar{m}   (k_A-k_B) \big) }
	{  \big(4 k_I^2 +  (k_A-k_B)^2\big) \bar{m} + 2 \big(2 k_I + k_A+k_B \big)   } \, ,
\end{eqnarray}
\begin{eqnarray}
	\beta_B &=& \frac{2  \big(1 - \bar{m}   (k_B-k_A) \big) }
	{  \big(4 k_I^2 +  (k_A-k_B)^2\big) \bar{m} + 2 \big(2 k_I + k_A+k_B \big)   } \, .
\end{eqnarray}

\subsection{Heat flow in the small-$k_I$ regime}
\label{app:smallkI}

Since we are interested in 
the regime of  very small  $k_I$, we can suppress the contributions of $\sigma_{mA} \sim O(k_I^2)$ and $\sigma_{mB} \sim O(k_I^2)$, since $\sigma_A \sim O(k_I^0)$ and $\sigma_B \sim O(k_I^0)$, leading to
\begin{eqnarray}
J_1 & \simeq & \kappa_0\Big[\beta_A\, g_A(\sigma_A T_A)   +  
   \beta_B \,  g_B( \sigma_B T_B)  \Big] (T_A - T_B) \,.
\end{eqnarray}

Below, we write the explicit expressions of all the coefficients for three especial cases, where the asymmetry relies either on the spring constant, the mass or damping parameter.

\subsubsection{Different stiffness}
Assuming $m_A = m_B = m$ and $\gamma_A = \gamma_B = \gamma$:
\begin{eqnarray}
      \kappa_0 &=& \frac{2\gamma k_I^2 }{ (k_A - k_B)^2 m + 2 (k_A + k_B)\gamma^2} \, , \\ 
      \sigma_{A} &=& \frac{1}{k_A}, \\
      \sigma_{B} &=& \frac{1}{k_B}, \\
      \beta_{A} &=& \frac{2  \big( (k_A - k_B)m + \gamma^2 \big) }{  (k_A - k_B)^2m + 2(k_A + k_B)  \gamma^2  }, \\
      \beta_{B} &=& \frac{2\big( (k_B - k_A)m + \gamma^2 \big) }{  (k_A - k_B)^2 m  + 2(k_A + k_B) \gamma^2    } .
\end{eqnarray}

\subsubsection{Different damping}
Assuming $m_A = m_B = m$ and $k_A = k_B = k$:
\begin{eqnarray}
      \kappa_0 &=& \frac{k_I^2}{k (\gamma_A + \gamma_B)} \\
      \sigma_{A} &=& \sigma_{B} = \frac{1}{k}, \\
      \beta_{A} &=& \frac{ \gamma_B}{k(\gamma_A + \gamma_B)} , \\
      \beta_{B} &=& \frac{ \gamma_A}{k(\gamma_A + \gamma_B)}.
\end{eqnarray}

\subsubsection{Different mass}
Assuming $\gamma_A = \gamma_B = \gamma$ and $k_A = k_B = k$:
\begin{eqnarray}
      \kappa_0 &=& \frac{ \gamma (m_A + m_B) k_I^2 }{k \big( k (m_A - m_B)^2 + 2 (m_A + m_B) \gamma^2 \big)}\, , \\ 
      \sigma_{A} &=& \sigma_{B} =  \frac{1}{k}, \\
      \beta_{A} &=& \frac{  2k(m_B - m_A)m_B + (m_A + m_B) \gamma^2  }{k  \big( k(m_A - m_B)^2 + 2 (m_A + m_B) \gamma^2 \big)  }, \\
      \beta_{B} &=& \frac{ \ 2k(m_A - m_B)m_A + (m_A + m_B) \gamma^2 }{k  \big( k(m_A - m_B)^2 + 2 (m_A + m_B) \gamma^2 \big)  }.
\end{eqnarray}

\end{document}